\documentclass{iaus}
\usepackage{graphicx}

\usepackage{xspace}
\newcommand{\Ha}{\ifmmode {\rm H}\alpha \else H$\alpha$\fi\xspace}
\newcommand{\Hb}{\ifmmode {\rm H}\beta \else H$\beta$\fi\xspace}
\newcommand{\nii}{\ifmmode [\rm{N}\,\textsc{ii}] \else [N\,{\sc ii}]\fi\xspace}
\newcommand{\oi}{\ifmmode [\rm{O}\,\textsc{i}] \else [O\,{\sc i}]\fi\xspace}
\newcommand{\oii}{\ifmmode [\rm{O}\,\textsc{ii}] \else [O\,{\sc ii}]\fi\xspace}
\newcommand{\oiii}{\ifmmode [\rm{O}\,\textsc{iii}] \else [O\,{\sc iii}]\fi\xspace}
\newcommand{\sii}{\ifmmode [\rm{S}\,\textsc{ii}] \else [S\,{\sc ii}]\fi\xspace}

\def\apj{ApJ}              
\def\mnras{MNRAS}          

\title[IAUS267. Emission line taxonomy and the nature of LINERs in the SDSS] 
{Emission line taxonomy and the nature of AGN-looking galaxies in the SDSS}

\author[Cid Fernandes et al.]   
{Roberto Cid Fernandes$^1$, 
Gra\.zyna Stasi\'nska $^2$, 
Natalia Vale Asari$^{1.5}$,
Ab\'{\i}lio Mateus$^{1}$, 
Marielli S. Schlickmann$^{1}$  
\and William Schoenell$^{1}$ ~~
(for the SEAGal collaboration)}
\affiliation{$^1$Departamento de F\'{\i}sica-CFM, Universidade Federal de Santa Catarina, Florian\'opolis, Brasil\\[\affilskip]$^2$LUTH, Observatoire de Paris-Meudon, France}
\pubyear{2010}
\volume{267}  
\pagerange{1--8}
\jname{Co-Evolution of Central Black Holes and Galaxies}
\editors{B.M.\ Peterson, R.S.\ Somerville, \& T.\ Storchi-Bergmann, eds.}
\begin{document}

\maketitle

\begin{abstract}
Massive spectroscopic surveys like the SDSS have revolutionized the
way we study AGN and their relations to the galaxies they live in.  A
first step in any such study is to define samples of different types
of AGN on the basis of emission line ratios. This deceivingly simple
step involves decisions on which classification scheme to use and data
quality censorship. Galaxies with weak emission lines are often left
aside or dealt with separately because one cannot fully classify them
onto the standard Star-Forming, Seyfert of LINER categories. This
contribution summarizes alternative classification schemes which
include this very numerous population. We then study how
star-formation histories and physical properties of the hosts vary
from class to class, and present compelling evidence that the emission
lines in the majority of LINER-like systems in the SDSS are not
powered by black-hole accretion. The data are fully consistent with
them being galaxies whose old stars provide all the ionizing power
needed to explain their line ratios and luminosities. Such retired
galaxies deserve a place in the emission line taxonomy.
\end{abstract}

\firstsection 
\section{Introduction}

The way things are named plays an important role in the organization
of scientific data.  When a galaxy is described as {\em
``Star-Forming''} (SF), one is lead to believe that star-formation is
the dominant (or maybe the only) source of energy.  When someone says
a galaxy is a {\em ``LINER''}, you are driven to think of a low
luminosity AGN, powered by accretion (possibly in a radiatively
inefficient regime) onto a super-massive black hole. Likewise, the word
{\em ``Seyfert''} works like an adjective to which one associates a
vigorous AGN, a dusty torus which blocks our view of the nucleus, a
bright and highly ionized Narrow Line Region, etc. Depending on the
reader's age range, memories of all those talks and papers about NGC
1068 will come to mind. Similarly, {\em ``SF $+$ AGN composite''} brings
to mind beasts like Mrk 477 (Heckman et al.\ 1997) or NGC 5135
(Gonz\'alez Delgado et al.\ 2001), where star formation and AGN with
comparable powers coexist.

In the context of in-depth studies of individual sources (like those
presented in the contributions by Steiner and Storchi-Bergmann in this
volume), such denominations are irrelevant formalities which play
little (if any) role in the interpretation of the data. On the
contrary, in the context of statistical studies of massive samples,
where one trades quality and detail by quantity, {\em taxonomy} plays
a key part. This paper, as many others in this volume and in the
current literature, deals with this kind of data. The link with the
central theme of this meeting, the co-evolution of central black holes
and galaxies, is that in order use survey-quality data to study such a
complex issue one must first make sure that black hole activity can be
correctly identified and quantified.

The way one identifies AGN in massive optical spectroscopic samples
like the SDSS is through their emission lines. Newcomers may have the
impression that all that could be said and done about spectral
classification of galaxies with emission lines has already been said
and done. We first remind the reader that an awful lot of emission
line galaxies (ELGs) in the SDSS simply cannot be reliably classified
using standard classification schemes because some of the required
lines (specially \Hb, but also \oiii) are just too weak (\S2).  After
reviewing current classification schemes (\S3), we then present
alternative diagnostic diagrams (and corresponding equations for class
division boundaries) which allow placing this large population of Weak
Line Galaxies (WLGs) within SF, Seyfert and LINER classes (\S4). Cid
Fernandes et al.\ (2010, CF10) opens up details on this revised
taxonomy. Finally (\S5 \& \S6), we take advantage of our detailed
stellar population and emission line analysis of the whole SDSS DR7
(soon available at www.starlight.ufsc.br) to make the bridge between
emission line classification and physical properties of the host. This
leads to a surprising interpretation as to the nature of SDSS
LINERs\ldots

\section{Weak Line Galaxies: Examples and the size of the problem}

\begin{figure}
\begin{center}
 \includegraphics[bb= 80 175 560 495,width=0.99\textwidth]{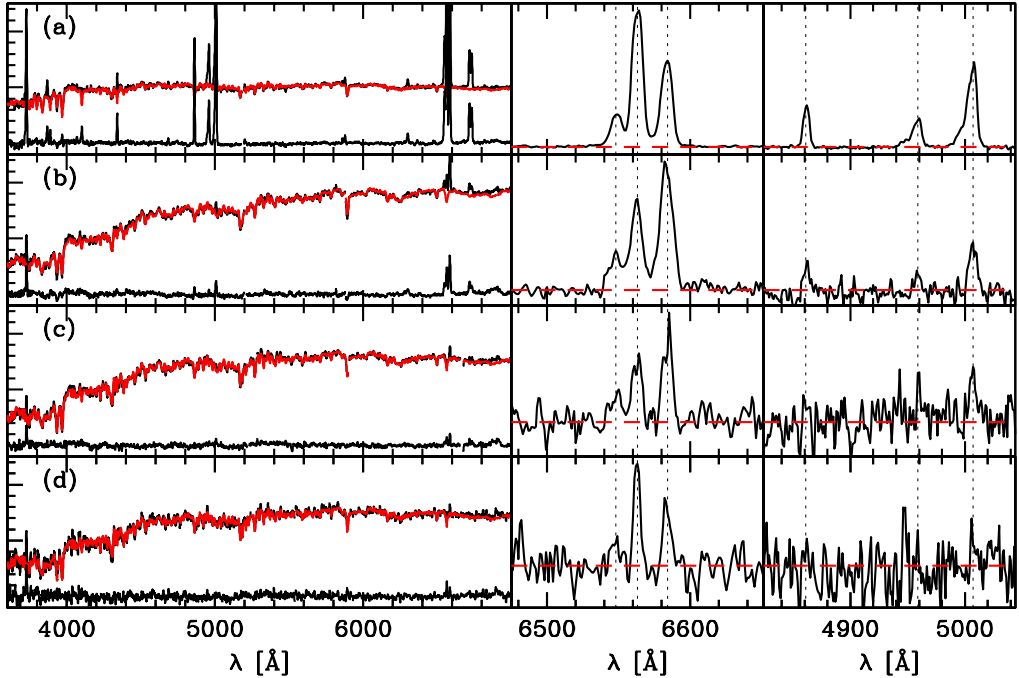}
\caption{Example SDSS spectra, including our {\sc starlight} fits and
the residual spectra. The right panels are zooms of the pure emission
spectra around \Ha and \Hb (the dashed horizontal line marks the zero
flux level). The top two are easily classifiable, but the bottom ones
are not.}
\label{fig:ExampleSpectra}
\end{center}
\end{figure}

Fig.\ \ref{fig:ExampleSpectra} shows four SDSS spectra. The top one is
the spectral classifier's dream. Its emission lines are so strong that
one could tell it is a Seyfert 2 from miles away. Example b has weaker
lines, but still strong enough for an unambiguous classification. From
its \Hb, \oiii, \Ha and \nii fluxes, all of which are detected at 3
sigma or better confidence, one can confidently put this galaxy in the
LINER's bin. Things get tougher in example c, where \oiii, \Ha and
\nii are all detected at $S/N \ge 3$, but \Hb is not. Example d is
even worse, as neither \Hb nor \oiii have decent detections.

Close to 80\% of galaxies in the SDSS have \Ha and \nii lines detected
at $S/N \ge 3$, but $\sim 1/3$ of these have less convincing
detections of either or both of \Hb and \oiii.  Cases c and d in Fig.\
\ref{fig:ExampleSpectra} illustrate these WLGs. They are the spectral
classifier's nightmare. From the high \nii/\Ha one can be reasonably
certain that these are AGN-like systems, but there is no well
established method to diagnose whether these are Seyferts or LINERs.
As many as 2/3 of the sources with $\log \nii/\Ha > -0.2$ have weak
\Hb and/or \oiii.

Is there a way of rescuing this huge population from the
classification limbo? The answer is yes. To do that, we must go back
to the taxonomy drawing board.

\section{BPT-based emission line taxonomy}

\begin{figure}
\begin{center}
 \includegraphics[bb= 40 170 500 590,width=0.99\textwidth]{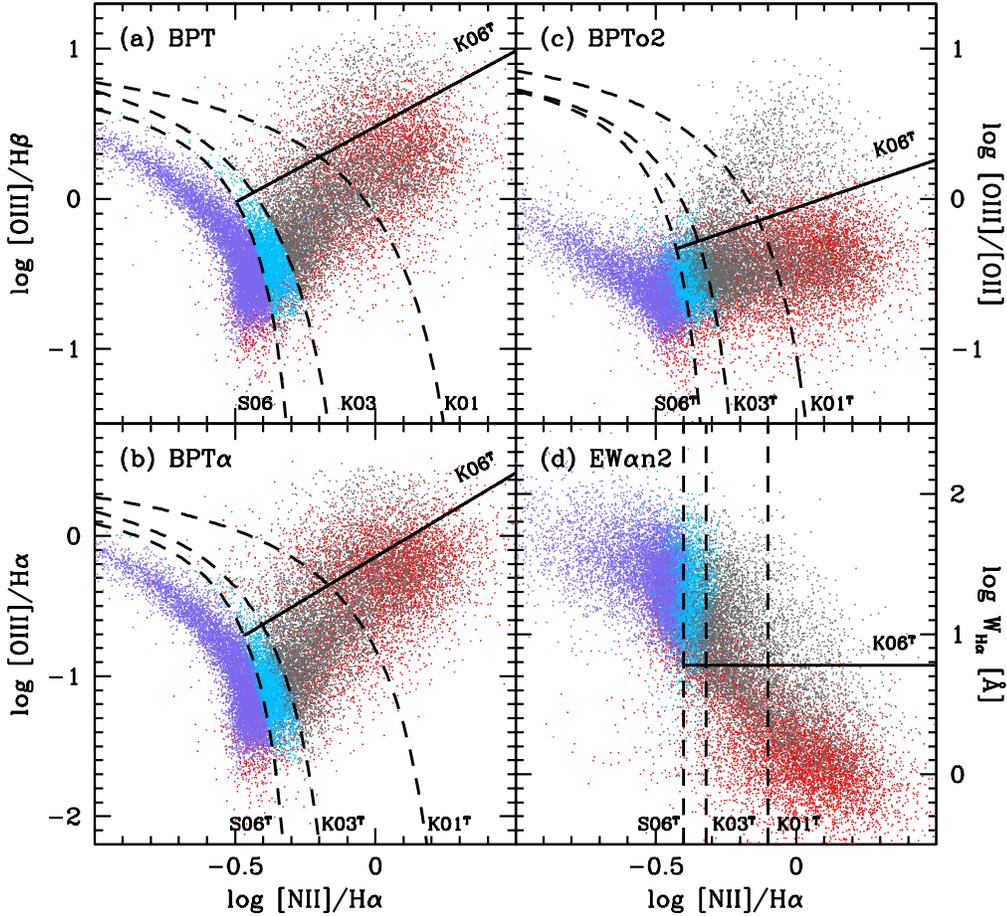}
\caption{The BPT (top right) and three alternative diagnostic
diagrams.  Dashed lines show three widely used SF/AGN division lines
(S06, K03 and K01), while the solid line divide Seyferts from LINERs
(K06). A superscript ``T'' is used to denote dividing lines which were
transposed from the originally defined (e.g., BPT-based) boundaries.
See Table \ref{tab1} for equations. Points in red have $S/N < 3$ in
either or both of \Hb and \oiii (WLGs).}
\label{fig:DivLines}
\end{center}
\end{figure}

According to the Oxford dictionary, taxonomy is the branch of science
concerned with classification.  The word finds its roots in the Greek,
taxis meaning arrangement/order, and nomia meaning distribution.  The
ultimate icon of order in the distribution of emission lines
properties of galaxies is the BPT diagram: $\oiii/\Hb \times \nii/\Ha$
(Fig.\ \ref{fig:DivLines}a). Its two well defined wings correspond to
SF galaxies (left wing), and systems where the ionization source is
harder than that produced by massive young stars (producing more
energetic photo-electrons and thus enhanced collisionally excited
lines). The right wing is commonly called the AGN wing, as this is
where bona fide AGN are located.

Up until not so long ago a mixture of art (ad hoc curves) and science
(grids of photoionization models) was used to draw class division
lines in diagnostic diagrams like the BPT. The task of classifying
ELGs on the basis of their emission line ratios was greatly simplified
with the statistics of the SDSS. With so many points to plot, the
morphology of diagnostic diagrams practically spells out class
division boundaries, so that all one has to do is to find a suitable
mathematical expression of this empirical result.

That is what Kauffmann et al.\ (2003, K03) did to separate SF galaxies
from AGN in the BPT diagram. Their division line is the current
standard SF/AGN classification scheme. Stasi\'nska et al.\ (2006, S06)
proposed a slightly different division line, based on photoionization
models designed to match the upper boundary of the SF wing. Kewley et
al.\ (2001, K01) proposed a model-based ``extreme starburst'' line
which, as seen in Fig.\ \ref{fig:DivLines}a, does not match the
observed morphology of the BPT at all. This line is nowadays widely
used to isolate ``pure AGN'', and, in combination with the K03 line,
to define ``SF $+$ AGN composite systems'', even though it was not
designed to do either thing.

The separation of Seyferts from LINERS was revisited by Kewley et al.\
(2006, K06), who identified a split of sources in the right wing into
upper (Seyfert) and lower (LINER) branches, clearly visible in
diagrams like \oiii/\Hb $\times$ \oi/\Ha or \sii/\Ha (also in the BPT,
albeit more blurred). K06 devised a classification system based on
\Hb, \oiii, \oi, \Ha, \nii and \sii lines which tracks this observed
bimodality.  To convert their Seyfert/LINER classification scheme to a
simpler one based exclusively on the BPT, a 2D version of the optimal
separator method was employed, maximizing completeness and reliability
fractions (see CF10). This leads to the solid line in Fig.\
\ref{fig:DivLines}a (see also Table 1).

We thus have 3 versions of SF/AGN division lines (S06, K03, K01, the
latter of which is not really adequate to separate SF from AGN), plus
the K06 Seyfert/LINER classification scheme transposed to a straight
line in the BPT plane. Obviously, all of this only applies when one
has reliable \Hb, \oiii, \Ha and \nii fluxes at hand. That is OK for
examples a and b in Fig.\ \ref{fig:ExampleSpectra}, but not for c, d
and the whole population of WLGs they represent. To include this
forgotten population of ELGs one needs to devise classification
schemes which are more economic in terms of data-quality (emission
line $S/N$) requirements.

\section{Emission line taxonomy revisited to include Weak Line Galaxies}

\begin{table}
\begin{center}
\caption{Class boundaries for spectral classification in various diagnostic diagrams}
\label{tab1}
\small
\begin{tabular}{|l|c|c|c|c||c|c|c|} \hline 
\multicolumn{2}{|c|}{$x = \log \frac{\nii}{\Ha}$} & 
\multicolumn{3}{|c||}{SF/AGN}  & 
Seyfert/LINER  &
\multicolumn{2}{|c|}{\%}  \\ \hline
Diagram &
$y$     &
S06 &
K03 &
K01 &
K06 & 
All$^1$ & 
AGN$^2$ \\ \hline
BPT         & $\log \frac{\oiii}{\Hb}$  &  $0.96 + \frac{0.29}{x + 0.20}$  &  $1.30 + \frac{0.61}{x - 0.05}$ &  $1.19 + \frac{0.61}{x - 0.47}$ &  $1.01 x + 0.48$ & 67 & 29  \\
 & & & & & & & \\
BPT$\alpha$ & $\log \frac{\oiii}{\Ha}$  &  $0.46 + \frac{0.29}{x + 0.22}$  &  $0.68 + \frac{0.49}{x + 0.03}$ &  $0.69 + \frac{0.57}{x - 0.38}$ &  $1.20 x - 0.15$ & 81 & 57  \\ 
 & & & & & & & \\
BPTo2       & $\log \frac{\oiii}{\oii}$ &  $1.06 + \frac{0.26}{x + 0.24}$  &  $1.10 + \frac{0.33}{x + 0.11}$ &  $1.25 + \frac{0.48}{x - 0.21}$ &  $0.64 x - 0.06$ & 77 & 50  \\
 & & & & & & & \\
EW$\alpha$n2& $\log W_{\Ha}$    &  $x = -0.40$               &  $x = -0.32$              &  $x = -0.10$              &  $W_{\Ha} = 6$ \AA & 100 & 100 \\ \hline
\end{tabular}
\end{center}
\vspace{1mm}
\scriptsize{
{\it Notes:}\\
$^1$ Fraction of all galaxies which have $S/N \ge 3$ in all lines involved. \\
$^2$ Fraction of all AGN-like galaxies ($\log \nii/\Ha > -0.2$) which have $S/N \ge 3$ in all lines involved.}
\end{table}

The main challenge to classify WLGs is to find a replacement for \Hb,
the weakest of the 4 BPT lines. \Ha and \oii provide suitable
alternatives. They are much less affected by low $S/N$, and the
\oiii/\Ha and \oiii/\oii ratios carry similar physical information
content as \oiii/\Hb (with both caveats and advantages). This leads us
to the BPT$\alpha$ and BPTo2 diagrams shown in Figs.\
\ref{fig:DivLines}b and c.  The S06$^T$, K03$^T$, K01$^T$, and K06$^T$
division lines in these more inclusive diagrams are optimal
transpositions of the original S06, K03, K01 SF/AGN and the K06
Seyfert/LINER classification schemes.  Equations for these lines are
given in Table 1, which also shows that whereas in the BPT 67\% of the
ELGs can be classified on the basis of $S/N \ge 3$ data, in the BPTo2
and BTP$\alpha$ this fraction increases to 77 and 81\%. The gain is
much higher considering right wing galaxies alone, for which these
diagrams allow one to classify about twice as many sources as the BPT.

A complete solution to the classification of WLGs requires replacing
\oiii by a stronger line with physically equivalent diagnostic power,
but there is no such a thing. A different way of looking at emission
lines is to combine line ratios and equivalent widths. It so happens
that replacing \oiii/\Hb by the equivalent width of \Ha provides an
efficient and very cheap classification scheme. In the EW$\alpha$n2
diagram (Fig.\ \ref{fig:DivLines}d), SF are separated from AGN by
$\nii/\Ha = -0.4$ for the S06 scheme and $-0.32$ for the K03 one,
while Seyferts and LINERs split at $W_{\Ha} = 6$ \AA. Inevitably, the
completeness and reliability fractions of these optimal transpositions
are not as good as for other diagrams, meaning that the
classifications obtained with this diagram do not match perfectly
those obtained with more standard ones. Yet, given its much larger
applicability, we dare suggesting the EW$\alpha$n2 diagram should
replace the BPT as a basis for spectral classification. Those
interested in AGN-host connections should welcome this proposition, as
an equivalent width provides a more direct metric of such connections
than a line flux ratio.

\section{Star Formation Histories across the BPT and EW$\alpha$n2 diagrams}

\begin{figure}
\begin{center}
 \includegraphics[bb= 30 170 560 680,width=0.99\textwidth]{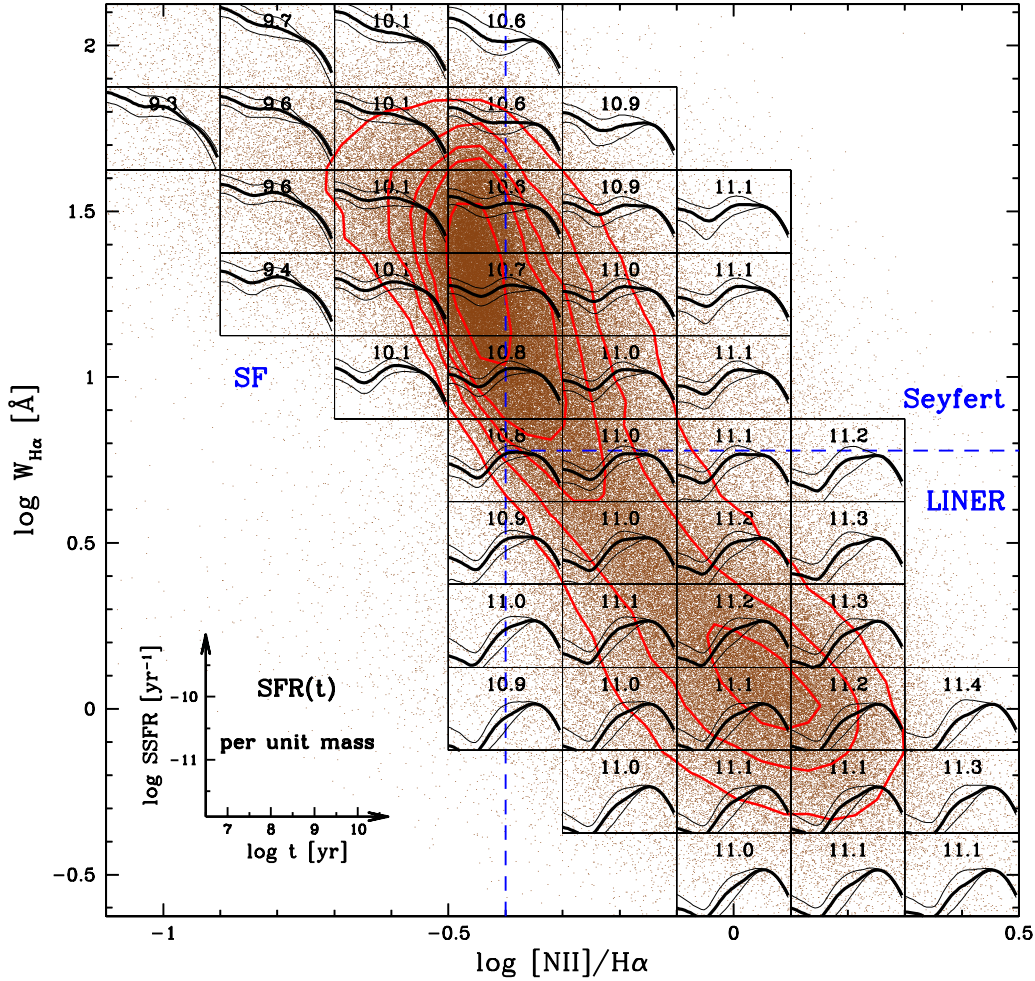}
\caption{SFHs across the EW$\alpha$n2 diagram. Points and contours
indicate the location of ELGs in the SDSS DR7. The diagrams is chopped
into boxes containing $> 500$ galaxies. The number at the top of each
box is the median logarithm of the stellar mass (in M$_\odot$). The
curves shows the $t$-by-$t$ median specific star-formation rate
against the lookback time, as show in the inset. Thinner lines tracer
the 16 and 84 percentiles. Dashed lines mark the transposed S06-SF/AGN
and K06-Seyfert/LINER division lines. Based on studies of globular
clusters, we warn that the small upturn in the SFHs of predominantly
old systems at very young ages ($< 10^7$ yr) may be a side effect of
blue horizontal branches not included in the stellar population
models.  }
\label{fig:SFHs_on_EWHaN2}
\end{center}
\end{figure}

\begin{figure}
\begin{center}
 \includegraphics[bb= 30 170 560 680,width=0.99\textwidth]{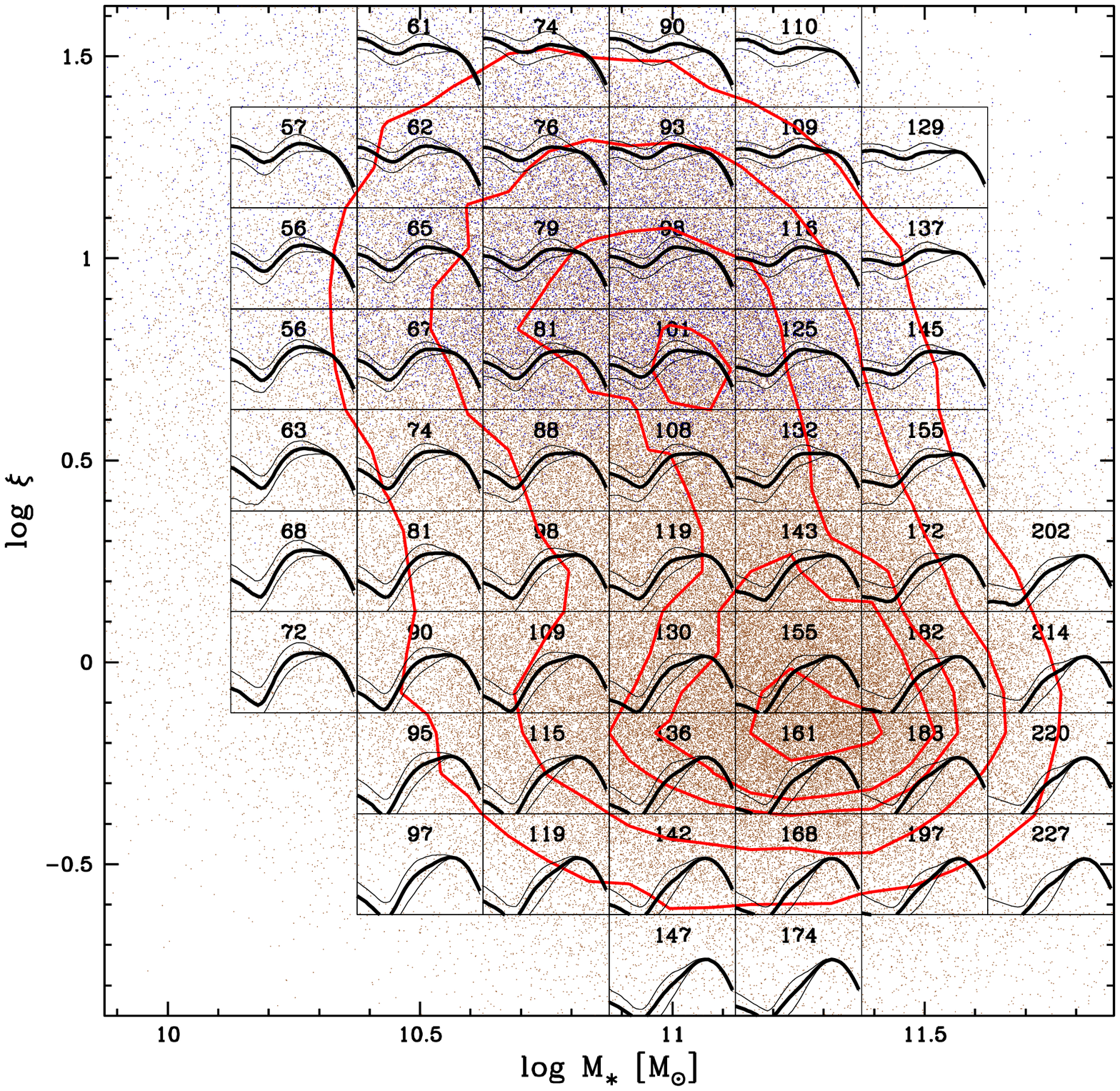}
\caption{SFHs across the $M_\star$ versus $\xi$ diagram for galaxies
classified as AGN according to the $\log \nii/\Ha > -0.40$ criterion
(S06$^T$ in Fig.\ \ref{fig:DivLines}d). (The SFH curves are on the
same scale as 
in Fig.\ \ref{fig:SFHs_on_EWHaN2}.) Sources
with $W_{\Ha} > 6$ \AA\ (Seyferts in the EW$\alpha$n2 diagram) are
painted in blue.  Numbers within each box represent the median
$\sigma_\star$ (in km$\,$s$^{-1}$), believed to be a tracer  
of black-hole mass.  Notice the bimodality in the $\xi$
distribution. $\xi \sim 1$ indicates galaxies whose \Ha luminosity can
be entirely powered by its old stars, as in the retired galaxy model
of S08. The SFH of galaxies in the $\xi \sim 1$ peak show that these
systems have indeed stopped forming stars long ago.}
\label{fig:SFHs_on_M_x_xi}
\end{center}
\end{figure}

So far we have talked only about emission line taxonomy. Yet, one
should not loose sight of the fact that classifying galaxies is not a
goal in itself, but just a means of organizing data in terms of
observables which (hopefully) bear correspondence with the underlying
physics, mapping different phenomena (or regimes of a same mechanism)
onto different classes. As shown in previous papers by the SEAGal
(Semi Empirical Analysis of Galaxies) collaboration, our {\sc
starlight}-SDSS database contains far more information than emission
line properties. Stellar masses ($M_\star$), velocity dispersions
($\sigma_\star$), stellar extinction, mean ages, stellar metallicities
and full time-dependent Star Formation Histories (SFHs) are among the
most interesting ones. A gold-mine to examine the link between
emission line classes and physical properties of host galaxies.

Stasi\'nska's contribution in this same volume shows the SFHs of
strong line galaxies ($S/N \ge 3$ in {\em all} 4 BPT lines) across the
BPT diagram. From top to bottom along the SF wing (and thus for
increasing nebular metallicity) SFHs change from systems which are
nowadays forming stars at a much higher pace than in the past to
galaxies which have kept $\sim$ constant rates throughout their lives
(Asari et al.\ 2007). Among AGN-looking galaxies, one sees that recent
($t < 10^8$ yr) SF activity decreases strongly as one goes from
Seyferts to LINERs.\footnote{We note in passing that galaxies in the
BPT zone usually associated to ``SF $+$ AGN composites'' do show
substantial ongoing SF, but so do galaxies in the upper side of the
right wing (the Seyfert zone), a region often associated with ``pure
AGN'' in the current literature.  The bottom line is that there is no
simple way of isolating truly composite systems on the basis of
emission line data alone.}

That plot is heavily biased, particularly in the right wing, where
$\sim 2$ in every 3 galaxies are left out because of bad \Hb and/or
\oiii data (Table 1). To include these WLGs, let us see how SFHs look
in our most inclusive diagnostic diagram: The EW$\alpha$n2. Overall,
Fig.\ \ref{fig:SFHs_on_EWHaN2} confirms the general result obtained
from the BPT, that Seyferts have substantial recent SF whereas LINERs
do not. The main difference is that, as a result of removing the
prejudice against WLGs, the population of LINERs is now much
larger. With the exception of massive metal rich SF systems which have
weak-\oiii, most WLGs are LINER-like systems, in the bottom-right part
of the EW$\alpha$n2 diagram. LINERs with strong lines are just the tip
of an iceberg.

\section{Retired galaxies $=$ fake AGN}

Plots like Fig.\ \ref{fig:SFHs_on_EWHaN2} can be made for any choice
of axis.  The abundant information provided by our stellar population
analysis allows one to look at the data from less observable-oriented
perspectives, so lets re-do that plot with more physically oriented
variables. An obvious choice for a physically interesting x-axis is
the stellar mass. Out of numerous options for the y-axis, for reasons
which will soon become clear, we chose a rather unconventional one:
The ratio of the observed \Ha luminosity to the predicted \Ha output
due to photoionization by {\em old stars} ($t > 10^8$ yr):

$$
\xi = \frac{L_{\Ha}^{\rm observed}}{L_{\Ha}^{\rm predicted}(t>10^8 \,{\rm yr})}
$$

\noindent The denominator comes from our {\sc starlight} analysis,
which (with the aid of evolutionary synthesis models) allows us to
predict the ionizing radiation field emanating from the old stars in a
galaxy. Using case B recombination coefficients and neglecting escape
of ionizing photons, one derives the predicted $L_{\Ha}$.  Notice that
this choice for an y-axis makes no statement about what powers the \Ha
emission. The normalization of $L_{\Ha}^{\rm observed}$ by
$L_{\Ha}^{\rm predicted}(t>10^8 {\rm yr})$ can be seen as just a
natural unit to measure the \Ha power.

Fig.\ \ref{fig:SFHs_on_M_x_xi} shows the $M_\star$ versus $\xi$
diagram.  This being an AGN symposium, the plot is restricted to
AGN-like systems, defined as those with $\log \nii/\Ha > -0.40$.  Of
the many things that could be said about this plot, let us first
highlight the bimodality in the ``AGN'' population, strongly
reminiscent of the Seyfert/LINER dichotomy identified by K06 in their
inspection of diagnostic diagrams.  Indeed, Seyferts live at the top
part of this diagram, while LINERs populate the $\xi < 0.5$ region,
peaking at $\xi$ slightly below 1.

$\xi = 1$ means that exactly all \Ha photons can be explained as
coming from photoionization by post-AGB stars and white dwarfs,
ionizing sources which are seldom considered relevant, specially among
AGNauts. It so happens that $\xi \sim 1$ is also the center of the low
peak in the bimodal distribution of sources in Fig.\
\ref{fig:SFHs_on_M_x_xi}. Looking at the SFHs of these galaxies one
sees that they have retired from forming stars long ago.  Given the
factor of $10^5$ difference in ionizing fluxes for young ($t < 10^7$
yr) and $> 10^8$ yr populations, any ongoing SF, even at small levels,
would move these galaxies to the $\xi \gg 1$ regime.  Similarly (but
on a more semantic vein), any AGN worth being called ``active'' should
be able to produce an ionizing field stronger than that produced by
the least powerful stellar populations. Hence, $\xi \sim 1$ galaxies
really ought to be retired, and their central black holes must be
fasting, otherwise $\xi$ would be higher.

{\underline {\it But can retired galaxies mimic AGN?}} If you haven't
done so yet, this is the time to read Stasi\'nska et al.\ (2008). The
self-consistent stellar population $+$ photoionization models
presented there show that retired galaxies can indeed mimic AGN in
terms of emission line ratios and luminosities, and that about 1/4 of
SDSS LINERs with $S/N \ge 3$ in all BPT lines can be explained in this
way.  Including WLGs, this fraction increases tremendously, to the
point that the Seyfert/LINER dichotomy does not seem to be a
manifestation of two regimes of black-hole accretion, but rather a
consequence of two entirely different phenomena: non-stellar versus
stellar ionization (ie, bona-fide AGN $\times$ retired galaxies, or
true $\times$ fake AGN).

LINERs have long been known to comprise a rather mixed bag, some with
unequivocal evidence of AGN (broad lines, variability, X-rays, etc.).
Our claim is that the emission lines in most objects called LINERs in
the SDSS are not powered by black-hole accretion, but by old
stars. Hence, beware of fake AGN!

\end{document}